\documentclass[aps,superscriptaddress,twocolumn,amsfonts,amsmath,showpacs,floatfix]{revtex4}

\usepackage{graphicx}
\usepackage{amssymb}  
\begin{document}

\title{Landscapes, dynamic heterogeneity and kinetic facilitation in a simple off-lattice model.}

\author{R. K. Bowles}
\email{richard.bowles@usask.ca}
\affiliation{Department of Chemistry, University of Saskatchewan, Saskatoon, SK, S7N 5C9, Canada}

\author{I. Saika-Voivod}
\affiliation{Department of Chemistry, University of Saskatchewan, Saskatoon, SK, S7N 5C9, Canada}

\begin{abstract}
We present a simple off-lattice hard-disc model that exhibits glassy dynamics.  The inherent structures are enumerated exactly, transitions between metabasins are well understood, and the particle configurations that act to facilitate dynamics are easily identified.  The model readily maps to a coarse grained dynamic facilitation description.

\end{abstract}

\date{\today}

\maketitle

 When a liquid is cooled fast enough to avoid crystallization, it becomes a supercooled fluid before eventually forming an amorphous solid. The feature of this transformation that is least understood is the lack of any structural signature associated with the dramatic slowing down of relaxation phenomena that occurs just prior to the temperature where the solid structure is frozen in. While there are many theoretical models describing glassy dynamics, three important concepts include the landscape paradigm, dynamic heterogeneity and kinetic facilitation.
 
 The potential energy landscape (PEL), which was originally introduced by Goldstein\cite{gold69} and later formalised by Stillinger and Weber\cite{stil82} in terms of inherent structures, describes the total system as a single point moving through the high-dimensional $N$-body potential energy function of the configurational coordinates. While the topography of this surface can be essentially characterized by local minima connected by saddle points, the exponential increase in the number of distinct minima with system size means that a complete description of the PEL is  only possible for small systems\cite{wal03}. Nevertheless, this approach has become an important theoretical and computational tool for the study of glasses because much of the interesting supercooled and glassy behavior can be connected to the statistical properties of the landscape. Some recent examples include the investigation of thermodynamic finite size effects\cite{heuer1};  the existence of an ideal thermodynamic glass transition at positive temperature\cite{bow00,deb03} and the connection between the configurational entropy, or number of minima accessible to the fluid, and structural relaxation\cite{spe}, via the Adam and Gibbs relation\cite{ag}.
 
 Experiments and simulations have shown that spatially heterogeneous dynamics is a general phenomenon of supercooled fluids nearing the glass transition\cite{ed00}. In particular, simulations show there are large domains of particles that move very slowly while other more mobile particles tend to cluster together in string-like arrangements\cite{harr95}. This raises an important question; How are these dynamic features connected to the configurational structure of the fluid? Studies of the {\it propensity} of a particle in a single configuration to move a long distance have confirmed the presence of a definite configurational component to dynamical heterogeneity\cite{harr04}, but they also show that this is not connected to any obvious structural indicators like free volume. 
 
A recent theoretical development\cite{gar02,gar03} that focuses on the important role of dynamic heterogeneity in glassy systems is based on dynamic facilitation\cite{fre84}. The central idea is that the dynamical structure observed in supercooled liquids is the consequence of local dynamical rules that restrict trajectory space, rather than any particular property of the static interaction potential. To exemplify this,  Garrahan and Chandler have used the Fredickson-Andersen\cite{fre84} and East\cite{jac91} spin lattice models, which only allow spins the opportunity to flip if certain local configurational constraints are statisfied, to examine some generic features like ``hopping time'' distribution functions\cite{gar03} and exchange times\cite{gar05} that might be measured in experiment. However, these coarse grained models avoid the question concerning the nature of the configurational component that gives rise to the local dynamical rules in structural glasses.

 In this letter we present a simple off-lattice model for which we can clearly identify the local configurational component and dynamical rules that give rise to dynamic heterogeneity and facilitation. We also obtain a complete description of the landscape for the model, including a complete enumeration of all the inherent structures, how they are simply connected and organised into metabasins, making this a system for which the main glassy phenomenologies, the potential energy landscape, dynamic heterogeneities and kinetic facilitation can be investigated at the same time. We explore how these approaches describe the glassy behavior of the model and how, in this simple case, all three are interconnected.
 
 The model consists of $N$ 2-dimensional hard discs of diameter $\sigma$ confined between two hard walls (lines) of length $L$  separated by a distance $H=(h+1)$ where lengths are given in units of $\sigma$. The particle-particle and particle-wall interaction potentials are given by 
\begin{equation}
\begin{array}{lcl}
V(r_{ij})=\left\{\begin{array}{ll}
0 &r_{ij}\geq\sigma\\
\infty &r_{ij}<\sigma
\end{array}
\right .
&:&
V_{w}(r_{i})=\left\{\begin{array}{ll}
0&r_{y}\le\left|h/2\right|\\
\infty&\mbox{otherwise}
\end{array}
\right .
\end{array}\mbox{,}\nonumber
\end{equation}
respectively, where $r_{ij}=|{\bf r_{j}}-{\bf r_{i}}|$ is the distance between particles and $r_{y}$ is the component of the position vector for a particle perpendicular to the wall. The occupied volume is then $z_{2d}=N\pi\sigma^{2}/(4L(h+1))$, but because of the quasi 1-dimensional nature of system, it is useful to use the occupied length $z=N\sigma/L=z_{2d}4(h+1)/\pi\sigma$. Finally, by restricting the channel diameter to $h<\sqrt{3/4}$ we ensure that only nearest neighbors can interact. This also prevents the particles from passing each other. We perform event driven molecular dynamics simulations on a system with $h=0.866$. Time $t$ has units of $\sigma(m/kT)^{1/2}$.

\begin{figure}[b]
\includegraphics[width=2.7in]{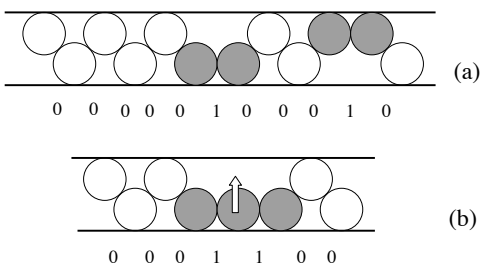}
\caption{(a) Stable packing configurations. The most dense arrangements are white. Particles in a defect appear grey. (b) The loose particle in a divacancy allows the system to unjam and eliminate the two defects. The ``0''s and ``1'' denote the bond assignments for the configurations. }
\label{pack}
\end{figure}

 To construct the hard particle equivalent to the potential energy landscape we need to obtain the distribution of jammed packed configurations with respect to their packing density\cite{spe,stil64}.
 Hard discs, in 2-dimensions are locally jammed if they have at least 3 contacts that are not all in the same semicircle. By restricting the channel radius to values of $h<\sqrt{3/4}$ we ensure particles can only interact with the two discs  on either side so a jammed particle has two disc-disc contacts and one disc-wall contact.  As a result, there are only two types of particle arrangement that lead to locally jammed particles (see Fig.~\ref{pack}a). 
The most dense arrangement requires the contacting neighbors to form a ``V"  with the central disc so that a series of densely packed discs forms a zig-zag structure. A defect, or locally less dense packing has two particles on the same side of the tube.  The defect behaves like a vacancy in that a particle in the defect can hop into the empty ``lattice" site but  it is not actually possible to add another particle to the system at the defect unless $h\ge\sqrt{3/4}$. Later, we will see that the presence of these defect sites plays an integral role in the glassy behavior of the fluid at high densities. The collectively jammed configurations of the entire system\cite{tor03}, or inherent structures are made up of combinations of these local packing arrangements, excluding those that contain neighboring defects because, as Fig.~\ref{pack}b shows, the central disc in a divacancy is free to move vertically which would allow the packing to unjam.  
 
 To count the number of packings with a density $z_{0}$ we note that, by drawing a ``bond'' between a particle and its neighbor on the right and writing a ``1" corresponding to a defect bond (i.e. when two discs are jammed together on the same side of the channel) and a ``0" otherwise, we can develop a lattice gas description of a jammed configuration, as shown in Fig.~\ref{pack}.  The total number of bonds must equal the number of particles $N$.  If $M$ is the number of defect bonds we can divide a configuration into blocks of  zeros and ones with $M_{01}$  $01$ boundaries. The total number of possible configurations is given by
\begin{equation}
\label{Ngall}
N_{g}(z_{0})=\frac{M!(N-M)!}{\left[M-(M_{01}/2)\right]!\left[N-M-(M_{01}/2)\right]!\left[(M_{01}/2)!\right]^{2}}\mbox{ .}\\
\end{equation}
If we allow all configurations, including unstable arrangements such as  ``11" and  ``111" etc, then Eq.~\ref{Ngall} becomes the ising model expression $N_{g}(z_{0})=N!/M!(N-M)!$.
To count only the stable packings, we remove all those configurations including neighboring defects, or ``11" arrangements, by enforcing $M_{01}=2M$ so that Eq.~\ref{Ngall} reduces to
\begin{equation}
\label{Ngstable}
N_{g}(z_{0})=\frac{(N-M)!}{M!(N-2M)!}\mbox{ .}\\
\end{equation}
The density of a jammed state is given by $z_{0}=\left[(1-\theta)\sqrt{1-h^{2}}+\theta\right]^{-1}$,
where $\theta=M/N$ is the mole fraction of defects, and the configurational entropy is
$S_{c}/Nk=(1/N)\ln N_{g}(z_{0})=(1-\theta)\ln(1-\theta)-\theta\ln\theta-(1-2\theta)\ln(1-2\theta)$.
The distribution of inherent structures is binomial with a single close packed structure with no defects and a single least packed structure with $\theta=0.5$ and maximum number of inherent structures occurring with $\theta=1/2-\sqrt{5}/10$. The same binomial distribution of inherent structures is found for a mixture of non-additive one-dimensional hard rods\cite{bow00}.

To complete the density landscape picture, we need the contribution to the entropy associated with the configurations that map to each inherent structure. The partition function of an individual basin could be calculated numerically using constrained simulations\cite{spe} but, for our immediate purposes, it is sufficient to make use of the simple double line model originally introduced by Wojciechowski et al\cite{wol81} which maps the model onto a mixture of 1-dimensional hard rods that interact with their neighbors with diameter $\sigma$ if both discs are on the same side of the channel or $\sigma(1-h^{2})$ if they are not. The resulting 1-dimensional approximation for the partition function of a basin with an inherent structure of density $z_{0}$ is $Q_{g}(z,z_{0})=(\Lambda^{N}L^{N}/N!)(z_{0}-z)/z_{0}$
where $\Lambda$ is the thermal DeBroglie wavelength of a disc. The equilibrium properties of the fluid are given by maximising the total entropy so that $S_{f}(z)/kN=\ln N_{g}(z_{0}^{*})+\ln Q_{g}(z,z_{0}^{*})$, where $z_{0}^{*}$ denotes the value of the limiting density that satisfies $\left(\partial S_{f}/\partial z_{0}\right)_{z}=0$.

\begin{figure}[b]
\includegraphics[width=2.7in]{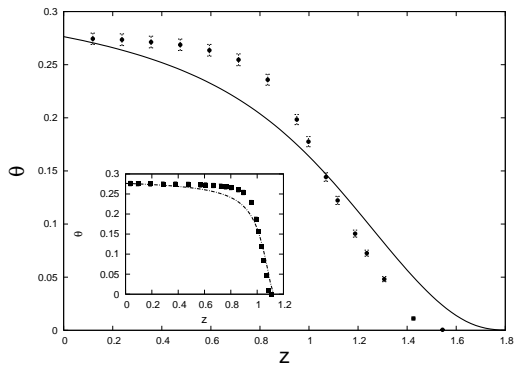}
\caption{$\theta$ measured from simulation (points) compared to the theory (lines) for a channel of width $h=0.866$, as a function of $z$ The insert: $h=0.5$. }
\label{ndef}
\end{figure}

 A key feature of the model is that the number and position of the defects indentify which basin the system is visiting. One method of counting the defect concentration in the fluid is to periodically interrupt the simulation and compress the configuration to its inherent structure. However, due to the nature of the model, it is actually possible to determine the lattice model description of the inherent structure from an instantaneous configuration by considering the relative positions of the neighboring particles and using the triangular constraint used by Speedy\cite{spe} to examine configurations of bulk disc systems. For example, if a particle is below the line drawn between the centres of its neighboring discs, the particle will pack against the  bottom wall, otherwise it becomes jammed against the upper wall. Fig.~\ref{ndef} shows the average concentration of defects obtained from a simulation of $N=5000$ particles as a function of density compared to that predicted by the model. At low densities, the system visits the set of basins that maximizes its configurational entropy but, at higher densities, the defects are eliminated in order to create more free volume. Due to the binomial distribution of inherent structures, the ideal glass transition for the model only occurs at close packing when the number of defects goes to zero. Some divacancies were also observed in the simulations.  For a system with $h=0.866$, at $z=0.11$, about 3.6\% of the defects appear in divacancies  and this concentration decreases below 0.5\% by $z=0.5$. The insert shows that the model improves as the channel narrows because the 1-dimensional partition function becomes a better approximation. 

Recent simulation studies of supercooled fluids have focused on relating structural relaxation to the organisation of the PEL into superstructures called metabasins. However, the notion of a metabasin is difficult  to define. The system is usually considered to have escaped from one metabasin to another during s simulation when a considerable change in the potential energy of the quenched structure correlates with changes in the particle positions\cite{buc00}. The relaxation of the system is then studied in terms of the distribution of waiting times for hops between metabasins\cite{dol03}. In the present model we can group all the basins belonging to inherent structures with the same number of defects into a single metabasin. The system moves between basins in the same metabasin when a particle hops to the vacant site in a defect. The transition between metabasins occurs through configurations containing a divacancy. Fig.~\ref{pack}b shows that the translation of the central disc results in the elimination of two defects. The creation of defects occurs via the same transition state.

 We can also directly identify the configurational elements that lead to dynamic heterogeneity and local rules for dynamic facilitation of defect hoping and defect elimination events. As the density of the fluid increases, the particles become caged by their neighbors and so they can only rattle around their local lattice site. However, the defects act as local regions of ``excitation'' and particles located in a defect can also hop to the vacant site. This is a very simple example of facilitation in which the local dynamics rules arise out of local packing constraints. As yet, we have not identified a configurational element that leads to the creation of defects.
 
\begin{figure}[t]
\includegraphics[width=3.0in]{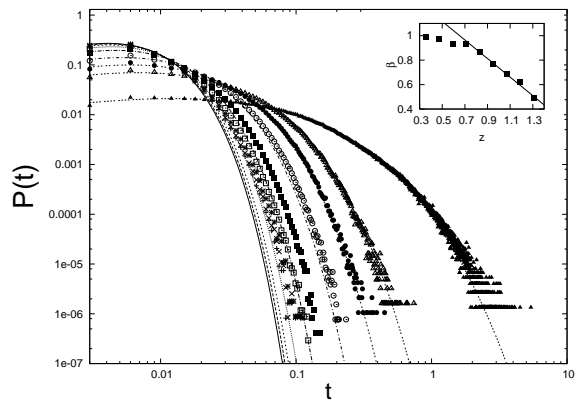}
\caption{Distribution of waiting times. From the right $z=1.306,1.188,1.069,0.950,0.831,0.712,0.593,0.475,0.356$. The lines represent best fits to the expression for $P(t)$ in the text. The inset shows the best fit parameter $\beta$ as a function of density. }
\label{wait}
\end{figure}

  Fig.~\ref{wait} shows the distribution of waiting times for transitions between metabasins on the PEL, i.e.~when the number of defects in the system changes. At high densities, where the caging of particles means that hopping at a defect is the only way to move between basins, we see long relaxation times that are associated with the need for defects to diffuse together and the distribution is well described by the facilitated dynamics model with
$P(t)\propto\left(t/t_{rel}\right)^\beta \exp\left[-(t/t_{rel})^{\beta}\right]$, where $t_{rel}$ and $\beta$ are both best fit parameters. The insert shows that the exponent for the stretched exponential becomes linearly dependent on density at high densities and in fact, the best fit line suggests that $\beta$ would go to zero at close packing. At low densities, where there are no barriers between basins on the landscape and no local packing constraints to require hopping between basins to be facilitated, the expression for $P(t)$ is no longer a good fit for the waiting times distribution. 

\begin{figure}[t]
\includegraphics[width=3.in]{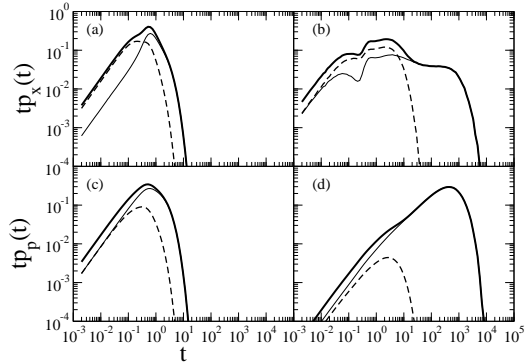}
\caption{Exchange times ($p_{x}(t)$) and Persistence times ($p_{p}(t)$) distribution functions. (a) $t p_{x}(t)$at $z=0.712$ (b)  $tp_{x}(t)$ at $z=1.425$.  (c) $tp_{p}(t)$ at $z=0.712$ (d) $tp_{p}(t)$ at $z=1.425$. Heavy lines denote the full distribution. Thin lines and dotted lines denote excitation ($0\rightarrow 1$)and de-excitation ($1\rightarrow 0$) events respectively. }
\label{cplot}
\end{figure}

The simplicity of our model allows us to examine some of the properties of the exchange and persistence times\cite{gar05} that help characterize heterogeneous dynamics. Fig.~\ref{cplot}
shows the distributions of exchange times, defined as the time between successive events (either excitation $0\rightarrow 1$ or a de-excitation $1\rightarrow 0$) at a particular position in the fluid, and the persistence times, which measure the time a local region takes before it is either excited or de-excited for the first time. We used systems of $N\approx100\theta^{-1}$. Breaking the distributions into their components shows that the slow relaxation at high densities is dominated by the presence of well packed regions waiting to be excited by the appearance of a defect which is rare and must diffuse through the system.

\begin{figure}[h]
\includegraphics[width=2.8in]{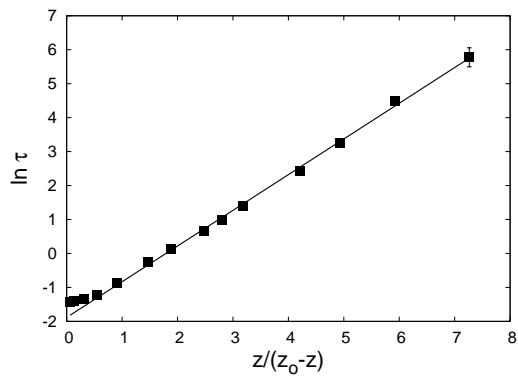}
\caption{Free volume plot for the average hopping times. }
\label{hoptimes}
\end{figure}

The exchange time distributions exhibit a rich structure at high densities that is associated with processes of defect creation, elimination and hopping that we will not describe in this letter. However, from the exchange times, we can extract an average hopping time associated with the time a defect takes to jump to its neighboring site by considering just the de-excitations but excluding any elimination events. Fig \ref{hoptimes} shows that the average hopping time for a defect  fits a free volume law, $\tau=A\exp[B z/(z_{0}-z)]$ above $z=0.8$ where $B\approx1$. This can be expected on physical grounds because, in order to hop from one site to the next, the cage must open sufficiently to allow the particle to move between its two neighbors.


In summary, we have presented a simple off-lattice model for which we can directly identify important configurational components and local dynamical rules, that lead to heterogeneous dynamics and facilitation, and are directly related to local packing constraints. This is consistent with the notion suggested by the propensity of particles in more complex systems\cite{harr04}, that dynamic heterogeneity arises from a configurational component. We have shown that our model displays glassy dynamics purely in terms of the landscape paradigm through its distribution of metabasin waiting times. 
However, we have also shown that the dynamics of our model can be described in terms of dynamic facilitation. 

We acknowledge NSERC for financial support.

 \end{document}